\documentclass[preprint2]{aastex63}

\submitjournal{ApJ}

\shorttitle{DR21 South Filament}
\shortauthors{Hu et al.}

\begin{document}

\newcommand\ammonia{NH$_3$}
\newcommand\pcmc{cm$^{-3}$} 
\newcommand\kms{km~s$^{-1}$}
\newcommand\kmsp{km~s$^{-1}$~pc$^{-1}$}
\newcommand\um{$\rm{\mu}$m}
\newcommand\trot{$T_{\rm rot}$}
\newcommand\tkin{$T_{\rm kin}$}
\newcommand\tdust{$T_{\rm dust}$}
\newcommand\msunpyr{M$_\odot$~yr$^{-1}$}
\newcommand\msunppc{M$_\odot$~pc$^{-1}$}
\newcommand\msun{\textit{M$_\odot$}}
\newcommand\lsun{\textit{L$_\odot$}}
\newcommand\pcmsq{cm$^{-2}$}
\newcommand\pcmcb{cm$^{-3}$}
\newcommand\mjypb{mJy~beam$^{-1}$}
\newcommand\herschel{\textit{Herschel}}
\newcommand\spitzer{\textit{Spitzer}}
\newcommand\water{H$_2$O}

\title{DR 21 South Filament: a Parsec-sized Dense Gas Accretion Flow 
onto the DR 21 Massive Young Cluster}

\correspondingauthor{Keping Qiu}
\email{kpqiu@nju.edu.cn}

\author[0000-0002-3286-5469]{Bo Hu}
\affil{School of Astronomy and Space Science, Nanjing University, 
163 Xianlin Avenue, Nanjing 210023, China}
\affil{Key Laboratory of Modern Astronomy and Astrophysics (Nanjing University),
Ministry of Education, Nanjing 210023, China}

\author[0000-0002-5093-5088]{Keping Qiu}
\affil{School of Astronomy and Space Science, Nanjing University, 
163 Xianlin Avenue, Nanjing 210023, China}
\affil{Key Laboratory of Modern Astronomy and Astrophysics (Nanjing University),
Ministry of Education, Nanjing 210023, China}

\author[0000-0002-6368-7570]{Yue Cao} 
\affil{School of Astronomy and Space Science, Nanjing University, 
163 Xianlin Avenue, Nanjing 210023, China}
\affil{Key Laboratory of Modern Astronomy and Astrophysics (Nanjing University),
Ministry of Education, Nanjing 210023, China}

\author[0000-0002-4774-2998]{Junhao Liu} 
\affil{School of Astronomy and Space Science, Nanjing University, 
163 Xianlin Avenue, Nanjing 210023, China}
\affil{Key Laboratory of Modern Astronomy and Astrophysics (Nanjing University),
Ministry of Education, Nanjing 210023, China}

\author[0000-0001-6630-0944]{Yuwei Wang}
\affil{School of Astronomy and Space Science, Nanjing University, 
163 Xianlin Avenue, Nanjing 210023, China}
\affil{Key Laboratory of Modern Astronomy and Astrophysics (Nanjing University),
Ministry of Education, Nanjing 210023, China}

\author{Guangxing Li}
\affil{South-Western Institute for Astronomy Research, Yunnan University,
Kunming, 560500 Yunnan, China}

\author{Zhiqiang Shen}
\affil{Shanghai Astronomical Observatory, 80 Nandan Road, 
Shanghai 200030, China}
\affil{Key Laboratory of Radio Astronomy, Chinese Academy of Sciences,
People's Republic of China}

\author{Juan Li}
\affil{Shanghai Astronomical Observatory, 80 Nandan Road, 
Shanghai 200030, China}

\author[0000-0001-6106-1171]{Junzhi Wang}
\affil{Shanghai Astronomical Observatory, 80 Nandan Road, 
Shanghai 200030, China}

\author{Bin Li}
\affil{Shanghai Astronomical Observatory, 80 Nandan Road, 
Shanghai 200030, China}

\author{Jian Dong}
\affil{Shanghai Astronomical Observatory, 80 Nandan Road, 
Shanghai 200030, China}

\begin{abstract}
    DR21 south filament (DR21SF) is a unique component of the giant
    network of filamentary molecular clouds in the north region of
    Cygnus X complex. Unlike the highly fragmented and star-forming
    active environment it resides, DR21SF exhibits a coherent
    profile in the column density map with very few star formation 
    signposts, even though the previously reported linear density of the 
    filament is an order of magnitude higher than the thermal stable 
    threshold.
    We derive the size (3.6~pc by 0.13~pc), temperature (10 to 15~K),
    and mass (1048~\msun) of DR21SF from Shanghai 65 m TianMa Radio 
    Telescope (TMRT) observations of \ammonia\ (1, 1) and (2, 2) 
    inversion lines in conjunction with the column density map from our 
    previous work.  Star-forming sites are identified along the 
    filament where gas temperature excesses.
    We find clear gradients in radial velocity and intrinsic line-width 
    along the spine of the filament.  The gradients can be well interpreted
    with a scenario of an accretion flow feeding DR 21 
    at a mass transfer rate of $1.1 \times 10^{-3}$~\msunpyr.
    Based on the analysis of its kinematic temperature, intrinsic 
    line-width and mass distribution, we
    conclude that DR21SF is in an overall trans-critical status, which
    indicates an early evolutionary stage.

\end{abstract}

\keywords{ISM: molecules --- ISM: structure --- radio lines: ISM}

\section{Introduction} \label{sec:intro}

  Understanding how stars, especially massive stars ($M > 8$~\msun),
  arise from the interstellar medium (ISM) has long been a major 
  challenge in modern astrophysics. Even though the relationship between
  filaments and star formation activities in molecular clouds
  has been proposed by both theoretical \citep[e.g.,][]{1964ApJ...140.1056O,
  1992ApJ...388..392I} and observational \citep[e.g.,][]{1979ApJS...41...87S}
  works for decades, it was the studies based on \herschel\ Space 
  Observatory \citep[e.g.,][]{2010A&A...518L.104M, 2010A&A...518L..95H} that 
  ultimately demonstrated the ubiquitous presence  
  of filaments in the star-forming activities.  
  Filamentary structures in molecular clouds 
  are found to fragment into nodes with scales from 0.01 to 0.1~pc
  in a gravitationally unstable status \citep{2010A&A...518L.102A}.
  These nodes are candidates 
  to form low-/intermediate-mass stars or star clusters 
  \citep[e.g.,][]{1999ApJ...510L..49J},
  while in some cases they can be as massive as several hundred 
  solar masses, and are capable of forming massive stars or star clusters 
  \citep[e.g.,][]{2010ApJ...719L.185J, 2018A&A...616A..78M}.
  There are also scenarios
  of a high-mass star forming region (HMSFR) connected to one
  \citep[the ``head-tail'' scenario][]{2002A&A...385..909T} or 
  more \citep[the ``hub-like'' system][]{2009ApJ...700.1609M} filaments.
  In these filaments, flows of gas that converge to the HMSFR 
  caused by longitudinally collapse may exist 
  \citep{2012ApJ...756...10L, 2013A&A...555A.112P, 
  2018ApJ...855....9L}.
  The accretion flow along the filament is proposed to be a more efficient 
  mechanism than an isotropic accretion to prolong the accretion time for 
  the high-mass star forming cores to gain their 
  mass \citep{2009ApJ...700.1609M}.

  The Cygnus X giant molecular cloud complex offers a large sample of star
  formation regions which show diversities in both physical conditions
  and evolutionary stages. Thus, it is known as 
  an excellent laboratory for studying the star formation process.
  Lying $1.50^{+0.08}_{-0.07}$ kpc \citep{2012A&A...539A..79R} away,
  the DR~21 ridge, consisting of DR21(OH) and DR~21, is 
  the most massive and star-forming active filament in the Cygnus X complex.
  In the northern part of the ridge, lies the DR21(OH), which is an extremely
  massive \citep[34000 \msun,][]{2010A&A...520A..49S} 
  star-forming region with intensive 
  maser activities \citep{1998ApJ...497..800K, 2008MNRAS.384..719H}.
  \citet{2010A&A...520A..49S} find out that three sub-filaments 
  (F1, F2, F3) are directly 
  linked to DR21(OH),  showing apparent velocity gradients along their 
  spines and suggesting material flows towards  
  DR21(OH) \citep[][hereafter, H12]{2012A&A...543L...3H}.
  To the south resides the relatively more evolved massive star forming 
  region DR~21.  It harbors a group 
  of luminous \ion{H}{2} regions \citep{2003ApJ...596..344C} and massive
  protostars traced by Class I and II methanol masers 
  \citep[e.g.,][]{1998ApJ...497..800K, 2012A&A...539A..79R}. 
  DR~21 is also well-known by being associated with the most 
  luminous (1800 \lsun) and massive ($>$3000 \msun) outflow in the Galaxy 
  \citep{1986MNRAS.220..203G, 2007MNRAS.374...29D}.  
  Two filamentary structures (referred as S filament and SW filament by H12)
  attached to DR~21 have been identified in the column 
  density map.  The more prominent one, S filament, is referred
  as DR~21 South Filament (DR21SF, see Figure~\ref{fig:overview}) 
  in this work, and extends
  about 8\arcmin\ directly from DR~21 to the south.   

  From a larger perspective, a network of dense gas filaments in the
  northern part of
  Cygnus X is associated with all the major massive star forming 
  regions (e.g., DR~22, DR~23, W75N, DR~21, DR~17) in the area
  \citep[][hereafter, Cao19]{2010A&A...520A..49S, 2019ApJS..241....1C}.
  This network hosts a number of filaments, most of which are overall 
  fragmented and showing rich signposts of star formation.  
  In contrast, DR21SF exhibits
  a coherent profile in the column density map, and only sparse star 
  formation candidates are found \citep[e.g.,][]{2014AJ....148...11K} 
  towards it.  H12 reported an average linear mass of 500 \msunppc 
  for this filament, which is an order of magnitude higher than the typical 
  thermal critical linear mass, 10 \msunppc\ \citep{2013A&A...553A.119A}.
  Based on observations towards several nearby filaments,
  \citet{2013A&A...553A.119A} concluded that super-critical filaments
  should fragment into cores and the subsequent local core collapse occurs 
  faster than global cloud collapse and thus super-critical filaments should
  be highly fragmented. However, DR21SF seems to be distinct from the above
  scenario.

  This filament has also
  been detected by \citet{2006A&A...458..855S} in $^{13}$CO 2$\rightarrow$1,
  with a radial velocity ranges from -6 \kms to -1 \kms. It shows
  roughly a radial velocity gradient from south to north in the channel map,
  which is often seen in the filaments that connecting to HMSFRs. 
  The possibility of analyzing the physical status of DR21SF is limited 
  by the spatial resolution of their data (130\arcsec).
  In order to understand the nature of DR21SF and its relationship with the
  DR~21, we perform spectral line observations of \ammonia\ towards the 
  filament,
  which will be analyzed in conjunction with our existing column density map 
  of the filament derived from the \herschel\ data.
  The \ammonia\ transitions ($J, K$) = (1, 1) and (2, 2) trace 
  gas with densities of $\sim 10^4$ \pcmcb\ \citep{2004tra..book.....R} 
  with a wide range of excitation temperatures \citep{1983ARA&A..21..239H}.  
  Unlike $^{13}$CO, which is not sensitive to gas with density larger than
  $10^3$ \pcmcb, \ammonia\ has 
  been commonly used as an exclusive probe of cold dense gas 
  in star-forming molecular clouds.   We therefore carry out 
  single-dish observations of \ammonia\ (1, 1) and (2, 2) transitions toward 
  DR21SF using Shanghai 65 m TianMa Radio Telescope (TMRT), trying to connect 
  the internal gas kinematics of the filament
  with its unique column density profile. 

  In Section~\ref{sec:observ}, we present technical details of the 
  observations and the data; The observational results and derived parameters
  of DR21SF are presented in Section~\ref{sec:result}; 
  In Section~\ref{sec:discussion}, 
  we will discuss the stability and kinematics of DR21SF and
  propose an accelerating in-fall gas flow scenario for the filament.

\section{Observations and Data} \label{sec:observ}
  \subsection{Dust Continuum Emission Data}
    In order to derive the mass distribution of DR21SF,
    we adopt the column density map of Cygnus X complex 
    from Cao19.  \herschel\ PACS (160~\um) and SPIRE (250, 
    320, and 500~\um) images are employed in the spectral 
    energy distribution 
    (SED) fitting to produce the column density map with an angular 
    resolution of 18\arcsec.  
    Detailed description about how the column density map is derived 
    can be found in Cao19. 
    One should notice that the SED fitting in Cao19 did not include
    the PACS 70~\um\ data since the emission at this wavelength arises from 
    warm dust and a single temperature model is no longer applicable.
    Given that the filament is overall quiescent and at low temperature,
    the estimate of the column density would not be significantly affected
    by neglecting the 70~\um\ data.

  \subsection{Ammonia Emission Line Observations}
    We carry out observations of \ammonia\ (1, 1) and (2, 2) lines
    towards DR21SF in both on-the-fly (OTF) and point by point On-Off modes 
    with Tianma 65~m telescope to obtain both a position-position-velocity 
    data cube covering the whole filament, and high resolution spectra of 
    targeted positions. 
    The total bandwidth of 500 MHz (OTF mode) and 187 MHz (On-Off mode) are 
    split into two sub-bands centered at 23694.4955 MHz and 23722.6333 MHz 
    with frequency resolutions of 30.517~kHz (OTF mode) and 5.722~kHz 
    (On-Off mode), respectively, corresponding to velocity resolutions
    of 0.39~\kms (OTF mode) and 0.07~\kms (on-off mode).  The FWHM beam-size 
    is about 48\arcsec.
    
    In the OTF observations,
    we map an area of 4.5$\arcmin$ by 9.0$\arcmin$ covering DR21SF
    on 25/11/2017, 07/02/2018 and 09/02/2018.  
    The average system temperature of the observations in three days
    is about 70~K.  The nominal integration time on each sampled point 
    is about 3~minutes 
    in total.  Only the main component of \ammonia\ (1, 1) 
    is detected in the OTF observations because the filament is overall 
    faint in \ammonia\ emission and the effective integration time is limited 
    for such a large OTF map. 
    The final data cube is then produced by
    combining all data with weights inversely
    proportional to their system temperatures, and re-griding the  
    image to a pixel size of 16$\arcsec$, approximately one third of the 
    beam-size.  

    The On-Off observations are made with a long integration time of 
    10 minutes towards each target position. We observe a total of 10 
    positions along the spine of DR21SF with spacings approximately identical 
    to the beam-size, to cover the entire filament.
    We name these ten positions as P01 to P10 from north to south.
    We fit all the observed \ammonia\ (1, 1) profiles with 
    a five-component model, whereas (2, 2) transition line profiles 
    with single 
    Gaussian model since we fail to detect the satellite lines of this
    transition.

\section{Result} \label{sec:result}

  \subsection{Morphology and Mass} \label{subsec:morphology+mass}

    \begin{figure*}
    \centering
    \epsscale{1.2}
    \plotone{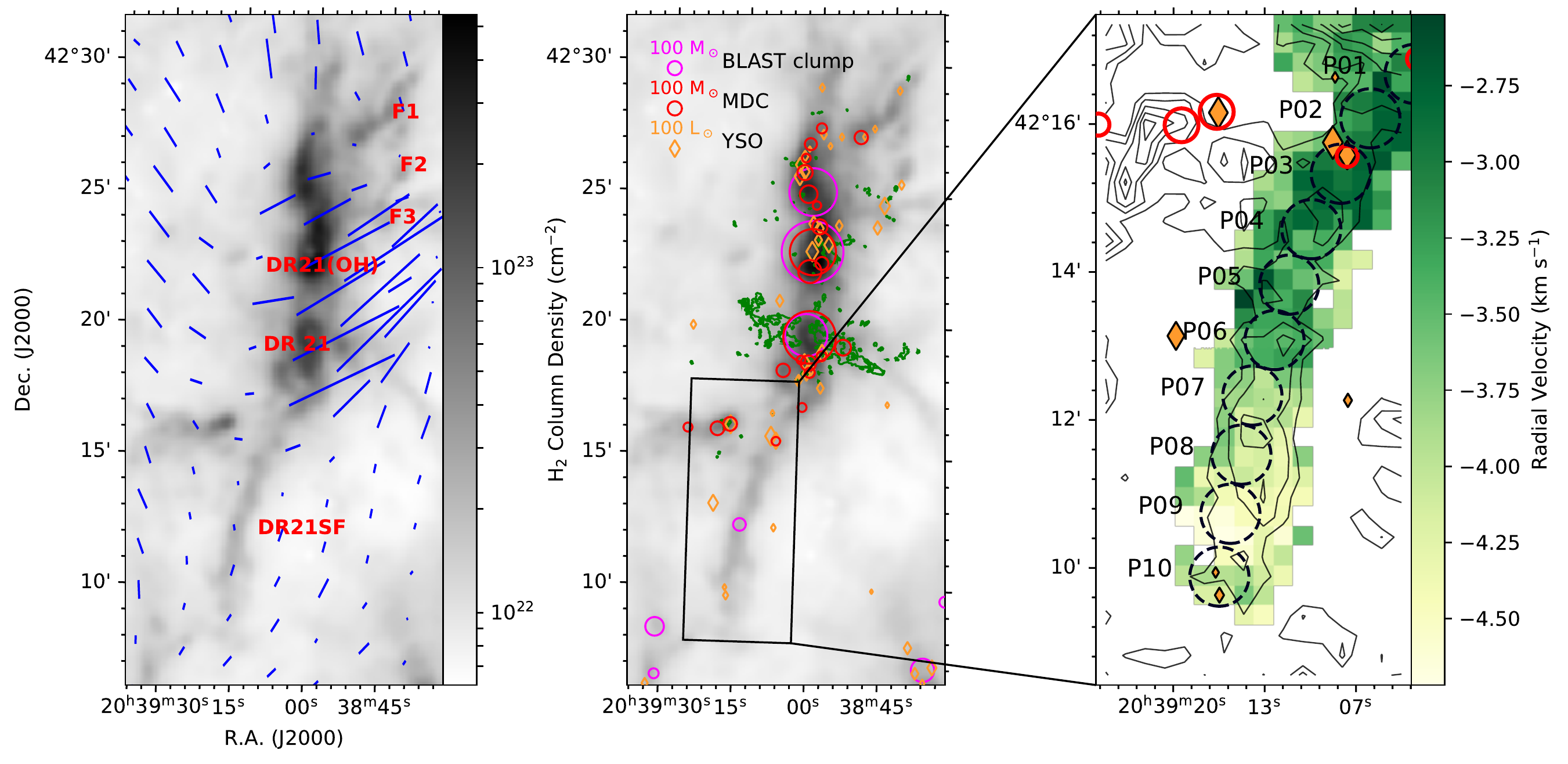}
    \caption{{\it Left panel:} The column density of the DR~21 area
    is shown by gray scale. 
    The orientation of the blue segments indicates the orientation of the 
    the magnetic field traced by polarized dust emission and length of
    the segments indicates the relative fraction of polarization.
    {\it Middle panel:} The gray scale is the same as the left panel.  
    Red and magenta circles denote
    the MDCs (massive dense cores) with mass greater than 35 \msun 
    recognized by Cao19 and
    clumps discovered by BLAST FIR observation,
    respectively, with the diameter proportional to their mass.  
    YSOs are denoted with orange diamond (open in left panel, 
    and filled in right
    panel), with the side length proportional to $\log L_{\rm YSO}$,
    where $L_{\rm YSO}$ is the luminosity of the YSO.
    Scales of MDC and BLAST clump with mass of 100 \msun and YSO with 
    luminosity of 100 \lsun are shown in the upper left corner.
    Outflows identified by 2.122 \um H$_2$ emission
    are marked with green contours.
    {\it Right panel:} The contour and the color map indicates the 
    moment-0 and moment-1 map of \ammonia\ (1, 1) 
    emission along DR21SF, respectively.  
    Black dashed circles indicate the beam-sizes and positions of 
    10 pointed observations. The MDCs, BLAST clumps, and YSOs shown
    in the middle panel are also 
    marked in this panel with corresponding marker and scale.
    \label{fig:overview}}
    \end{figure*}

\begin{figure}
    \plotone{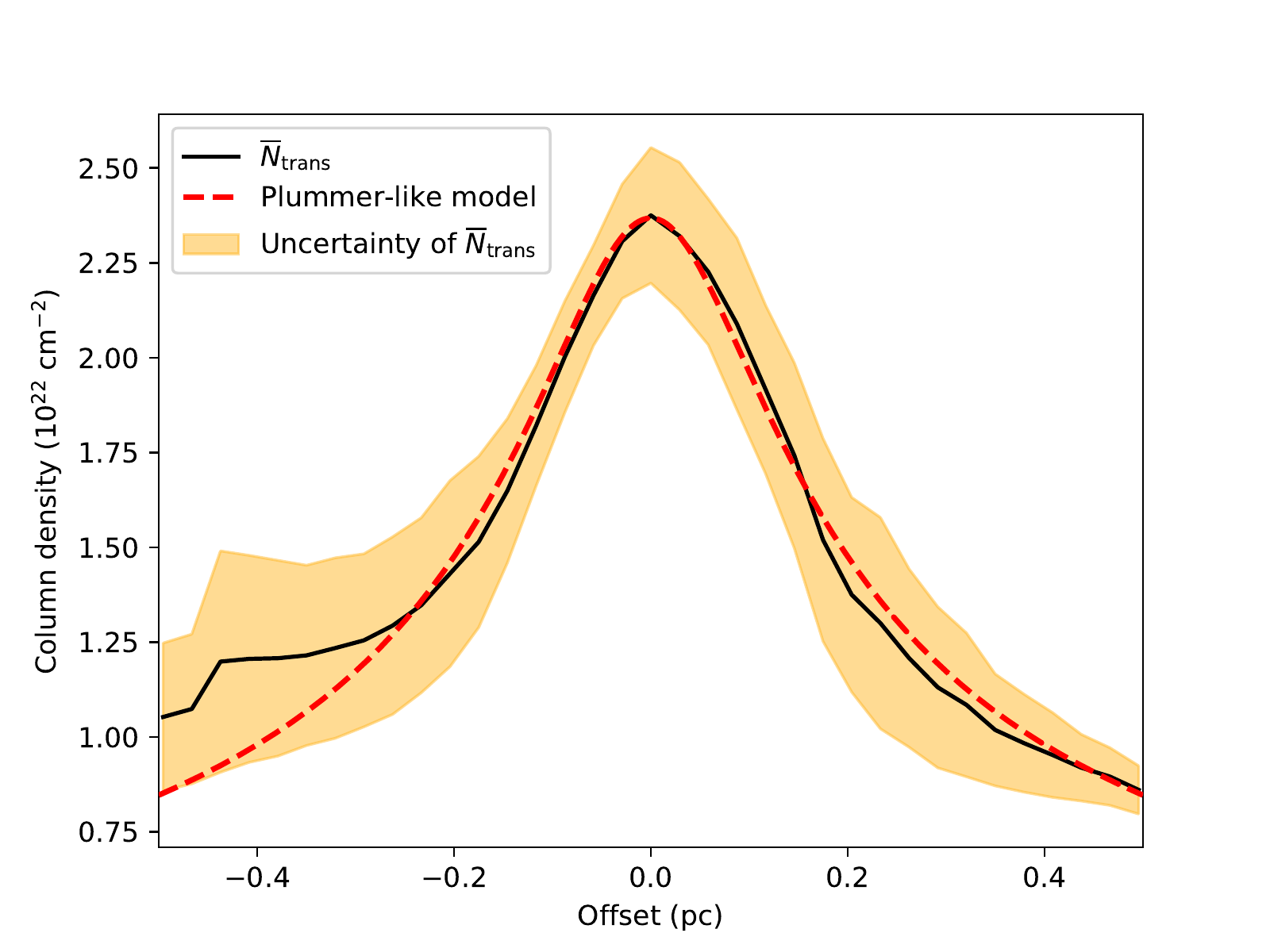}
    \caption{The averaged transverse column density 
        ($\overline{N}_{\rm trans}$) is calculated within a stripe of 1 pc
        wide long the spine of the filament on the column density map.
        $\overline{N}_{\rm trans}$ profile and its uncertainty are 
        denoted by the black line and the yellow belt, while the red
        dashed line represents the best-fit 
        Plummer-like model.
        \label{fig:width}}
    \end{figure}

    On the column density map, DR21SF is connected to DR 21 at its northern
    end, and extends $\sim$ 8\arcmin\ to the southeast.  
    Since DR21SF and DR~21 have similar radial velocities
    (about $-3$~\kms), it is reasonable to infer 
    that these two structures are physically associated 
    with each other. We therefore
    adopt trigonometric parallax distance of $1.50^{+0.08}_{-0.07}$ kpc 
    of DR~21 \citep{2012A&A...539A..79R} for DR21SF, which leads to a 
    length of 3.6~pc, 
    given an inclination angle of 18\degr\ (see discussion in 
    Section~\ref{subsec:scenario}).
    The width of DR21SF is derived by fitting the averaged column density
    profile perpendicular to the spine of the filament.
    We only employee the transverse column density profiles at 
    P03 to P09
    for the width estimation, since the column densities in the middle 
    segment of the filament are consistent.
    The asymmetry of the averaged transverse column density profile 
    (see Figure~\ref{fig:width}) is
    caused by the clump DR21-10 (identified by Cao19), which is on the east
    side of DR21SF and about 0.7 pc away from the spine of the filament.
    According to the cylinder model of an infinite gas filament
    \citep[e.g.,][]{1964ApJ...140.1056O, 2011A&A...529L...6A}, 
    whose radial volume density follows
    \begin{equation}
    \rho(r) = \frac{\rho_{\rm c}}{[1+(r/r_0)^2]^{p/2}} 
    \end{equation}
    the transverse column density can be described as a Plummer-like function:
    \begin{equation}
    N(r) = A_{\rm p}\frac{\rho_{\rm c} \sqrt{8} r_{\rm 0}}
           {(1+\frac{r^2}{8r_0^2})^{(p-1)/2}},
    \end{equation}
    where $A_{\rm p} = 
    \frac{1}{\cos{\theta}}\int\frac{{\rm d}u}{(1+u^2)^{p/2}}$,
    $\theta$ is the inclination and set to 18\degr,
    $\rho_{\rm c}$ is the central volume density of the filament,
    $r_0$ is the scale radius representing the flat distribution
    of the density at the spine of the filament. We fit the average
    transverse column density $\overline{N}_{\rm trans}$ profile with this 
    model. Shown in the lower
    panel of Figure~\ref{fig:width}, the best fit, which gives a minimum 
    $\chi^2$, results in $p = 1.7\pm0.1$, 
    $\rho_{\rm c} = (1.6\pm0.2) \times 10^4$~\pcmcb\ and 
    $r_0 = (4.2\pm0.6)\times 10^{-2}$~pc.  The errors are of one-sigma
    uncertainties.  The derived $p$ value is in agreement with that found 
    in filaments such as IC~5146 \citep[$1.5 < p < 2.5$,][]{2011A&A...529L...6A}
    and Serpens South filament \citep[$p =1.9\pm0.2$,][]{2013ApJ...766..115K}.  
    The central density $\rho_{\rm c}$ of DR21SF is significantly
    lower than that of several other
    star-forming filaments \citep[see Table 3 of][]{2018A&A...620A..62G}.
    When $p\approx2$ the FWHM of the radial column density profile of the
    gas cylinder is about $3 \times r_{0}$ \citep{2011A&A...529L...6A}, thus
    we estimate the width of DR21SF as 0.13 pc. This result seems in good
    agreement with the 0.1 pc ``universal'' width of filaments in molecular 
    clouds in the solar neighborhood proposed by \citet{2011A&A...529L...6A}.
    However, the resolution of the column density map we used is 18\arcsec,
    which corresponds to 0.13 pc at a distance of 1.5 kpc. We think the width 
    estimation is limited by this resolution, even though the
    fitting is performanced on the column density profile, whose FWHM is
    significantly larger than the resolution.

    The mass of DR21SF can be estimated from the column density map by
    \begin{equation}
    M = \sum N_i \mu m_H A_i,
    \end{equation}
    where $N_i$ is the column density of the $i$-th pixel on column density 
    map within the area of the filament, $\mu$ is the average molecular 
    weight, $m_H$ is the mass of a hydrogen atom, and $A_i$ is the area of 
    the $i$-th pixel at the distance of the source.
    We calculate the total mass within an area of a stripe
    along the spine of the filament with a width of 0.75~pc, which is
    the total width of the best-fit transverse column density profile above 
    the typical background level ($1 \times 10^{22}$\pcmsq).
    With $\mu = 2.8$, a total mass of 1048\msun\ is derived.
    Cao19 estimates the average relative uncertainty of column density to be
    about 40\% in the area of DR21SF. Considering the distance uncertainty
    is about 5\%, we claim that the uncertainty of the total mass is 50\%.
    
    The mass we estimated is smaller than 
    that (1350 \msun) proposed by H12, even though their result is within
    the uncertainty of ours. The discrepancy mainly comes from
    the systematic difference in the column density maps
    produced by H12 and Cao19.
    As a comparison, the column densities of the densest region in DR~21 
    and the nearby least dense regions reported by Cao19 are about
    $1.5 \times 10^{23}$ \pcmsq\ and $7.0 \times 10^{21}$ \pcmsq, 
    respectively, whereas H12 presents corresponding values of 
    $>1.0 \times 10^{24}$ \pcmsq\ and $\sim 1.0 \times 10^{22}$ \pcmsq,
    respectively. 
    The systematic difference can be explained by their data reduction
    approach. The SED fitting performed by Cao19 is based on images of 
    160 \um, 250 \um, 350 \um, 
    and 500 \um, whereas H12 does not include the data of 500 \um.
    The long wavelength data is essential to the SED fitting of cold 
    gas. For example, with a temperature of 15~K, the SED
    peaks at a wavelength of 340 \um.
    Thus, data at 500 \um\ will certainly help improve the 
    SED fitting of the cold gas in DR21SF. 
    The discrepancy should effect both the mass and the 
    temperature measurements as they are direct results from the SED fitting.  
    We do find a systematic difference about 3~K in the dust temperature
    of DR21SF between H12 and Cao19 as well.
    A column density about $1.2 \times 10^{23}$ \pcmsq\ was 
    independently derived from the archive data of 
    European Space Agency's Infrared Space Observatory (ISO) Long-Wavelength
    Spectrometer (LWS) by \citet{2007A&A...461..999J}
    towards the densest region in DR 21. This value is in good agreement
    with that of Cao19.
    We therefore conclude that the column density map presented by  Cao19 is 
    a better estimate compared to that by H12, therefore is adopted in this 
    work for the mass calculation.
    We also noticed that both of these two SED fitting processes did not
    take into account the 70 \um\ data. The dust emission at 70 \um\ mainly
    traces the warm gas component in the molecular clouds thus may introduce
    extra uncertainty to the SED if only a single temperature component
    of cold gas is applied in the fitting. Neglecting the warm gas
    component traced by shorter wavelength emission will certainly 
    underestimate the total mass of molecular clouds, especially for
    molecular clouds with high star formation rates. However, DR21SF
    shows very few star formation signposts (see discussion in 
    Section~\ref{subsec:stability+sf} below), therefore, we think
    this effect is neglectable compared 
    to other sources of uncertainty in the estimation of the total mass.
    
  \subsection{Optical Depth and Gas Temperature} \label{subsec:odepth+temp}

    \begin{figure}
    \plotone{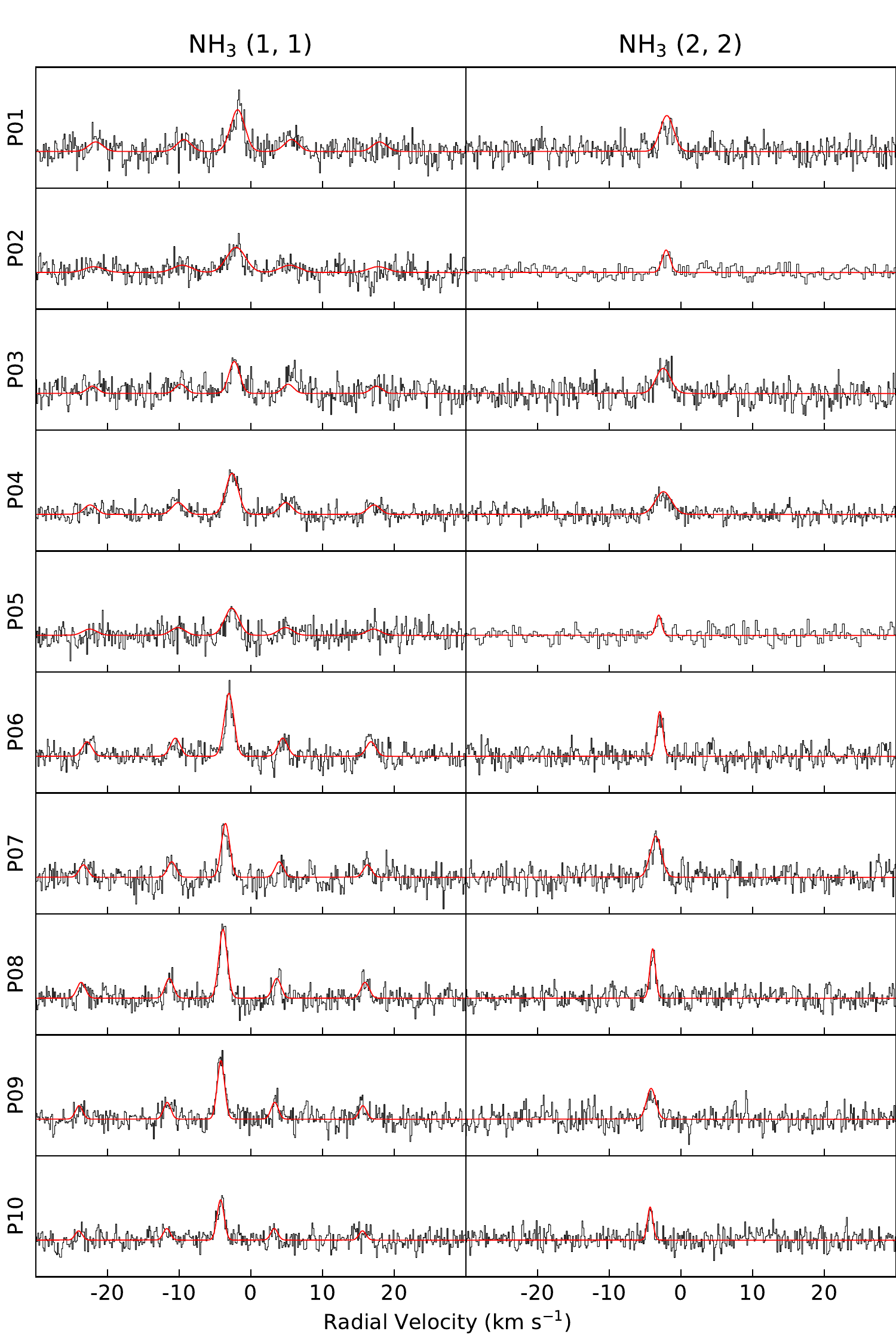}
    \caption{From top to bottom are the \ammonia\ (1, 1) (left column) and 
        (2, 2) (right column) line profile of P01 to P10. Black 
        lines indicate the data, whereas the red lines illustrate the
        best fit models. 
        \label{fig:line}}
    \end{figure}

    Data from the On-Off observations towards P01 to P10 are used to derive the 
    optical depth and gas temperature of the DR21SF.
    In Figure~\ref{fig:line}, the line profiles of both \ammonia\ (1, 1) 
    and (2, 2) transitions at P01 to P10 are presented. 
    To improve the SNR, we perform Hanning smoothing to the original data, and
    it results in an effective velocity resolution of 0.15 \kms.  With 
    assumptions of equal beam-filling factor and equal excitation temperature 
    for each hyper-fine transition, and a cosmic background 
    temperature of 2.7~K, we use the CLASS 
    package of the GILAS\footnote{\url{https://www.iram.fr/IRAMFR/GILDAS/}} 
    software to fit the \ammonia\ (1, 1) line profiles.  
    We only perform simple Gaussian fitting to the \ammonia\ (2,2) lines, 
    since we only detect the main component of this transition.
    Derived parameters are listed in Table~\ref{tab:p}.
    We follow Equation 4 in \citet{1983ARA&A..21..239H}
    to calculate rotational temperature (\trot). Since 
    \ammonia\ (1, 1) and (2, 2) transitions are not sensitive to temperatures
    significantly higher than 30~K \citep[e.g.,][]{2014ApJ...790...84L}, 
    the \trot\ $\gtrsim$ 30~K derived
    with this method can only be treated as a lower limit. 
    Based on \trot\ we are able to 
    calculate the kinematic temperature (\tkin), by applying the
    empirical relationship introduced by \citet{1983A&A...122..164W}:
    \begin{equation}
        \frac{1}{T_{\rm rot}} = \frac{1}{T_{\rm kin}}+\frac{1}{41.5}
        \ln[1+0.6\exp{-\frac{15.7}{T_{\rm kin}}}].
    \end{equation}
    Both \tkin\ and \trot\ for each position are also given in 
    Table~\ref{tab:p}.

    \begin{deluxetable*}{crrrrccccr}
    \tablecaption{Physical properties of \ammonia\ along the DR 21 south 
    filament.}
    \tablehead{
        \colhead{Position} & \colhead{R.A.} & \colhead{Dec.} 
        & \colhead{\trot} & \colhead{\tkin} & \colhead{$\tau(1, 1)$} 
        & \colhead{$V_{\rm lsr}$} & \colhead{$\Delta V_{\rm fit}$} 
        & \colhead{$\Delta V_{\rm int}$}
        & \colhead{$\mathcal{M}_{1D}$}
        \\
        \colhead{} & \colhead{($\degr$)} & \colhead{($\degr$)} & \colhead{(K)} 
        & \colhead{(K)} 
        & \colhead{} & \colhead{(\kms)} & \colhead{(\kms)} 
        & \colhead{(\kms)}
        & \colhead{}
    }
    \startdata
    P01 & 309.760 & 42.278 &   10.9 &    11.3 & 0.11 & -1.85 $\pm$0.01& 1.98 $\pm$ 0.14 & 1.86 & 10.18\\
    P02 & 309.773 & 42.268 &   21.4 &    25.0 & 0.10 & -2.07 $\pm$0.13& 2.63 $\pm$ 0.37 & 2.57 &  9.45\\
    P03 & 309.782 & 42.255 &$>$30.4 & $>$40.6 & 0.10 & -2.31 $\pm$0.12& 1.81 $\pm$ 0.42 & 1.68 &  4.83\\
    P04 & 309.791 & 42.243 &   12.1 &    12.7 & 0.10 & -2.63 $\pm$0.07& 1.99 $\pm$ 0.16 & 1.87 &  9.66\\
    P05 & 309.798 & 42.230 &   10.4 &    10.8 & 0.10 & -2.68 $\pm$0.10& 1.34 $\pm$ 0.27 & 1.20 &  6.71\\
    P06 & 309.803 & 42.218 &    9.7 &    10.0 & 0.10 & -3.05 $\pm$0.03& 1.17 $\pm$ 0.08 & 1.02 &  5.93\\
    P07 & 309.809 & 42.206 &$>$38.3 & $>$58.8 & 0.10 & -3.55 $\pm$0.03& 1.56 $\pm$ 0.12 & 1.42 &  3.39\\
    P08 & 309.813 & 42.192 &   22.1 &    26.1 & 0.10 & -3.89 $\pm$0.03& 1.02 $\pm$ 0.09 & 0.83 &  2.96\\
    P09 & 309.816 & 42.179 &   14.2 &    15.2 & 0.10 & -4.18 $\pm$0.04& 0.98 $\pm$ 0.10 & 0.78 &  3.66\\
    P10 & 309.819 & 42.165 &   22.3 &    26.3 & 0.10 & -4.23 $\pm$0.15& 1.13 $\pm$ 0.20 & 0.97 &  3.46  
    \enddata
    \tablecomments{The right ascension and declination are of J2000. 
    \trot\ and \tkin\ are the rotational and kinematic temperature, respectively.
    $\tau(1, 1, m)$ denotes the optical depth of the \ammonia\ (1, 1) transition.
    The $\Delta V_{\rm fit}$ and $\Delta V_{\rm int}$ stand for the fitted
    (``blend'') and intrinsic line-width, and the one dimensional Mach
    is listed in the last column.
    \label{tab:p}
    }
    \end{deluxetable*}

  \subsection{Velocity and Line-width} \label{subsec:velo+linewidth}
   
\begin{figure}
    \plotone{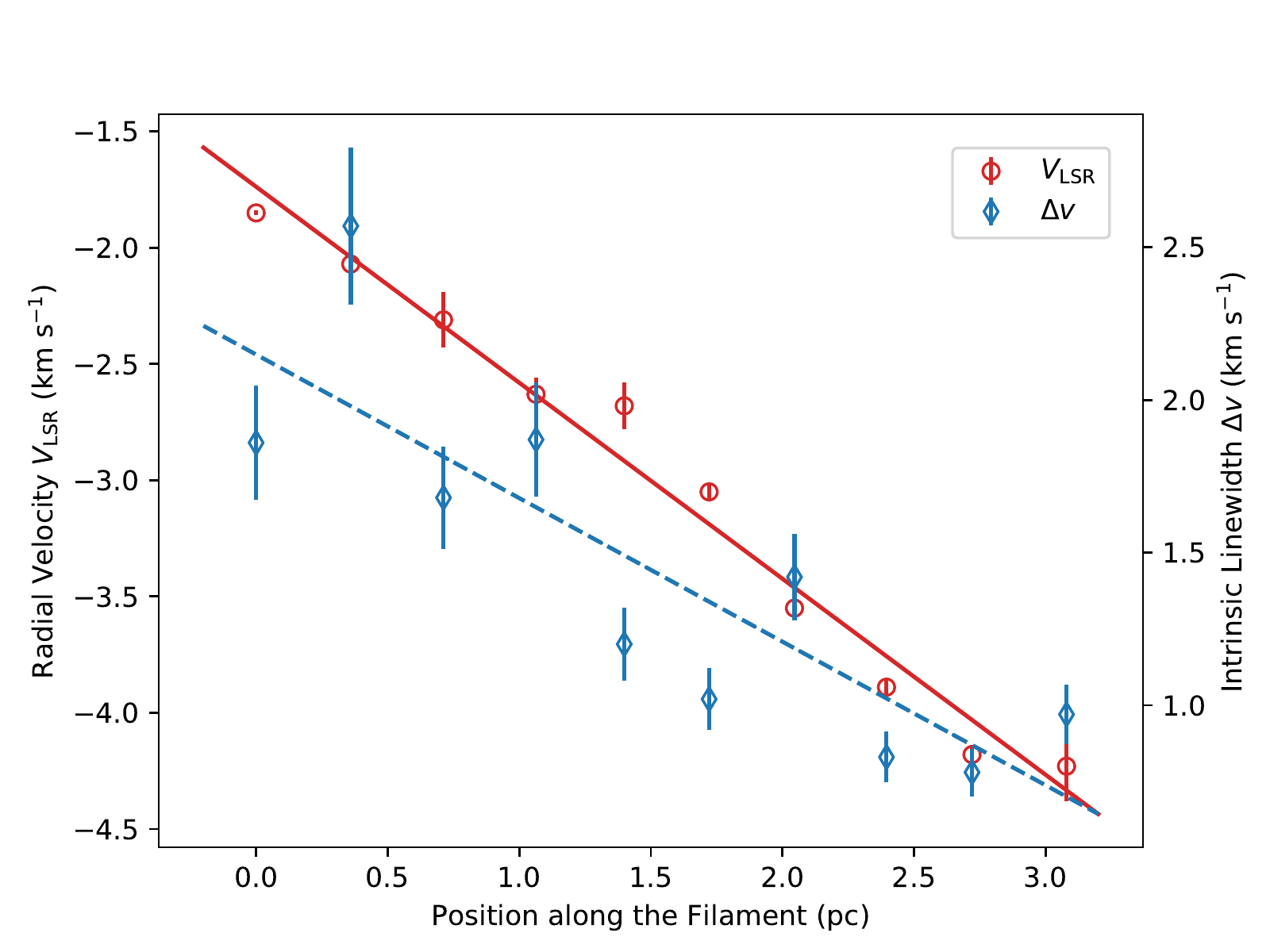}
    \caption{The red circles and solid line indicate the data and 
        the best-fit linear model of radial velocity, while blue diamonds 
        and dashed line indicate the intrinsic line-width and its best-fit 
        linear model.
    \label{fig:gradient}}
    \end{figure}

    The radial velocities along DR21SF with respect to local standard of 
    rest (LSR) range from 
    -4.23 to -1.85~\kms.  In the right panel of 
    Figure~\ref{fig:overview}, we present the first moment map of 
    the filament.  From both the first moment map   
    and the radial velocities presented in Table~\ref{tab:p} one can 
    easily see a gradient in radial velocity along the filament. 
    In Figure~\ref{fig:gradient}, we plot the radial velocity 
    as a function of the position along the filament with respect of P01. 
    The radial velocities are well characterized by the linear model, 
    which gives a gradient of $(0.8\pm0.1)$~\kmsp. 
    This value is comparable to 
    that in other similar filamentary molecular clouds, 
    e.g., 0.7 \kmsp\ in Orion filament \citep{1987ApJ...312L..45B}, 
    0.15-0.6 \kmsp\ in SDC13 \citep{2014A&A...561A..83P},
    0.8-2.3 \kmsp\ in DR21(OH) \citep{2010A&A...520A..49S}. 
    Moreover, a decreasing can also be seen in 
    the intrinsic line-width along the filament from north to south.
    We again apply a linear model to give a first order estimate
    of such a gradient. The best-fit result reads $(0.5\pm0.1)$~\kmsp.

    Due to the limited spectral resolution,
    the line-width directly derived from the observed line profile
    is the true line-width of the emission (intrinsic line-width) 
    convolved with an instrumental spectral resolution and blurred
    with adjacent hyper-fine structures.
    Also, the radial velocity gradient along the filament will broaden 
    the measured line-width when observed with a certain beam-size.
    To eliminate these effects, 
    we first subtract the instrumental spectral resolution and
    contribution of velocity gradient within the beam-size (0.3~\kms in 
    our case) quadratically from the measured line-width to derive a 
    ``blurred'' line-with.  We then follow the empirical relation
    \citep{1998ApJ...504..207B} between the intrinsic and the ``blurred'' 
    line-width to estimate the former. Results are listed in 
    Table~\ref{tab:p}. We evaluate the uncertainty of 
    intrinsic line-width to be 12\%, since the average one-sigma 
    uncertainty of fitted line-width is about 12\% and other uncertainties 
    are negligible.

\section{Discussion} \label{sec:discussion}
  
  Filamentary molecular clouds and their relationship with star formation
  have been intensively studied over the past decades.
  Detailed case studies have revealed physical
  properties of a number of remarkable star-forming filaments
  such as Orion A Integral Shaped Filament \citep{2019MNRAS.489.4771G,
  2018MNRAS.475..121S}, Nessie filament \citep{2018A&A...616A..78M},
  SDC13 \citep{2018A&A...613A..11W, 2016A&A...594A.118M}.
  These filaments exhibit complex gas kinematics, and often appear to 
  be highly fragmented along their major axes.
  On the other hand, tenuous filaments are often found to be connected 
  with large angles or orthogonal to the massive and star-forming filaments.
  For example H12 reported several filamentary streamers
  converge to the DR21(OH) with their column densities about a few
  orders of magnitude lower than that of the DR~21 ridge.
  Being a part of the giant filamentary network,
  DR21SF has a length about 4~pc, comparable with that of massive star-forming 
  filaments, but it is significantly less active in star formation.
  While, DR21SF is apparently thicker and more massive than the minor 
  and tenuous gas flows connected to the DR21(OH) ridge (sub-filament F1, F2 
  and F3 for example, see Table 1 of H12 for a comparison).

  So, what is the nature of DR21SF?  Is it a stable structure that is capable  
  of forming stars in itself? What is the physical connection between
  DR21SF and the massive young stars in DR 21?
  These questions are of great interests to our understanding of the star 
  formation activity in Cygnus X and the dynamical evolution of dense 
  molecular cloud filaments.

  \subsection{Star Formation and Stability} \label{subsec:stability+sf}

    Thanks to the deep and unbiased infrared surveys towards Cygnus X
    carried out with Balloon-borne Large Aperture Sub-millimeter Telescope
    \citep[BLAST,][]{2011ApJ...727..114R}, \herschel\ Space Telescope and 
    \spitzer\ Space Telescope, we are able to search deep embedded 
    young stellar objects (YSOs) and compact clumps which can potentially 
    form stars along the filament.
    We find signposts of star formation at P03, P07 and P10,
    in the surveys mentioned above. 
    As shown in Figure~\ref{fig:overview}, two compact mid-infrared (MIR) 
    objects are detected at P03 by 
    the \spitzer\ Space Telescope, and later recognized as Class I low-mass 
    YSO candidates by SED fitting using \spitzer, UKIDSS and 2MASS data
    \citep{2014AJ....148...11K}. 
    With luminosities about 505 \lsun and 86 \lsun,
    they are found to be coincident with
    an ``IR-quiet''\footnote{Cao19 defined ``IR-quiet'' MDCs as those
    with mid-IR flux lower than that of a B3-type stellar embryo, indicating
    the absence of high mass star formation in those cores.} 
    massive dense core (MDC, named as DR21-22 by Cao19) with a mass of 
    36.5~\msun.  The MDC shows no evidence of
    high mass star formation according to its IR luminosity. the two 
    objects are likely to be clusters of or individual 
    intermediate-/low-mass forming stars.
    At P07, a sub-mm object J203914+421221 is detected by 
    \citet{2011ApJ...727..114R} with BLAST.
    By the SED fitting of sub-mm and FIR data (500, 350, 
    250, 100, and~60 \um), the authors infer this object to be a 
    pre-stellar massive (80~\msun) clump at its very early stage with 
    a total luminosity of 270~\lsun\ and a temperature 20.2~K. 
    However, in a search of dense clumps
    with a mass completeness level of 35~\msun\ by Cao19, 
    no clump is detected at this position. 
    The inconsistency may come from distinct data 
    reduction procedures and spatial resolutions. 
    As argued by Cao19, the source extraction process is highly effected by
    the background diffuse emission removal, thus different 
    background removal algorithms may
    result in different detection. Also, the image resolution of 
    Cao19~(20\arcsec) is significantly higher than that of 
    \citet{2011ApJ...727..114R}~(1\arcmin), so the clump recognized by the 
    latter may be resolved into several less massive objects or relatively 
    extended structures.
    We also found two Class I YSOs with very low luminosities of 0.7~\lsun\ and 
    1.5~\lsun\ at P10. The less luminous one shows much more flatter SED with
    spectral index of $\alpha = 0.21$ than that of the other YSO $\alpha=1.67$. 
    \citet{1994ApJ...434..614G} referred such objects as ``flat spectrum'' 
    since their evolutionary stages show ambiguity between Class I and 
    Class II, and may statistically infer a later evolutionary 
    status than typical Class I YSOs. 
    \citet{2018ApJS..234....8M} carried out a 2.122 \um\ molecular hydrogen
    line survey of jets and outflows from young stars covering the whole 
    Cygnus X complex as part of the UWISH2 survey. The 2.122 \um\ molecular
    hydrogen emission line is excited by the shocks that produced by
    the high velocity outflow impacting the surrounding molecular gas.  
    We find no detection of 2.122 \um\ molecular hydrogen emission towards 
    all the YSOs identified in DR21SF in this survey. This is an
    indicator that the YSOs in DR21SF lack collimated jets,
    and may have only produced large angle outflows that are not able 
    to generate strong enough shocks in the natal clouds.

    The \ammonia\ observations (Table~\ref{tab:p}) show that
    the gas temperatures at P02, P03, P07, P08 and P10 are 
    significantly higher than typical temperatures (10 - 15~K) of cold 
    molecular clouds. 
    This can be explained by the objects we have found
    at P03, P07 and P10, as the on-going 
    star-forming process will heat up the natal clouds.
    It is also a direct proof of that the YSOs or clumps found in 
    these positions are physically associated,  
    rather than only visually coincident, with the filament.
    We find no YSO at P07, even though the gas 
    temperature and line-width are higher than the rest parts of the filament. 
    It is likely that the forming star or cluster
    is in its early evolutionary stage, 
    so it may not be visible as a near-/mid-IR point source.

    We then try to figure out that
    whether DR21SF is stable, or just a transient
    phenomenon, by investigating the stability using three time 
    scales: life time, crossing time, and free-fall time.
    In the case of DR21SF, the
    YSOs identified in the filament are Class I objects. 
    The age $\tau_{i}$ of such
    objects from pre-stellar to Class I stage is confined by 1~Myr 
    $\leq \tau_{i} \leq$ 2~Myr \citep{2005MNRAS.360.1506K, 2009ApJS..181..321E}.
    This will put a 1~Myr lower limit on the life time of the DR21SF.
    Recent observations towards many molecular cloud filaments 
    appear to favor a turbulence supported gas cylinder scenario. 
    At a temperature of 15~K, the thermal line-width is about 0.03 \kms for
    \ammonia, which is negligible compared to the measured 
    line-widths in DR21SF.  Therefore, we suggest that turbulence is the 
    dominant supporting force in the case of DR21SF.
    Considering the 3D velocity dispersion 
    $\sigma_{\rm 3D} = \sqrt{3}\sigma_{\rm aver} = 2.5 \pm 0.9$ \kms,
    where $\sigma_{\rm aver}$ is the average intrinsic line-width, the 
    crossing time then can be derived as 
    $t_{\rm cross} = R/\sigma_{\rm 3D} = 7.8 \times 10^4$ yr,
    in a turbulence supported case.
    The free-fall time of DR21SF can be estimated as
    $t_{\rm ff} = \sqrt{3\pi/32G\rho} = 1.3 \times 10^5$~yr, 
    where $\rho$ is the average volume density.
    Both $t_{\rm cross}$ and $t_{\rm ff}$ are about an order of magnitude
    smaller than the lower limit of the age of the filament. This indicates
    that DR21SF is a stable structure rather than a transient concentration
    of the ISM. However, the crossing time is significantly smaller than
    the free-fall time. This is a clear sign that the self-gravity of the 
    filament is not strong enough to overcome the turbulent pressure and 
    the filament would have been dispersed by the strong turbulence. 
    To solve this contradiction, we then have to look into the
    stability status of the filament.

    According to \citet{2000MNRAS.311...85F}
    the linear virial mass of turbulence supported filaments
    is given by 
    \begin{equation}
    (M/l)_{\rm vir} = 2\sigma^2/G = 
    84(\frac{\sigma_{\rm 3D}}{{\rm km s}^{-1}})^2 {\rm \msun pc^{-1}}.
    \end{equation}
    We calculate the virial mass using the equation above and the mass
    using the same method as the total mass estimation, for each of the 
    10 segments along the filament at P01 to P10.
    The virial parameter of each segment listed in Table~\ref{tab:vir}
    is the ratio of these two values. The virial parameter 
    along the filament is mostly around 1 to 2, suggesting that DR21SF
    is overall in a trans-critical status, that the gravity and
    turbulent pressure rival each other. At P02, 
    and P04, the virial parameters are significantly larger than 2.
    Since YSOs have already been identified at this area, the large 
    virial parameters do not really reflect stronger turbulence. 
    Instead, the
    feedback of the star formation activities, such as large angle
    outflows, may introduce perturbation to the surrounding molecular 
    clouds and increase the line-width. The large virial
    parameter at P07 and P10, comparing to their neighbor segments,
    can also be explained by the similar effect. This
    could be another evidence of the unseen star formation
    activities at extremely early stage at P07, independent from the
    temperature excess. We notice that the tail of filament 
    is in a supercritical status indicated by the small virial
    parameters. This is usually a reflection of effective collapse
    of the cloud under its self-gravity, and can possibly explain
    the later evolutionary stage of one of the YSOs identified
    at P10.
    
    \begin{deluxetable}{cccc}
    \tablecaption{The virial parameter of 10 segments along the
    DR21 south filament.}
    \label{tab:vir}
    \tablehead{
        \colhead{Segment} & \colhead{$M_{\rm vir}$} & \colhead{$M$}
        & \colhead{$\alpha$}
        \\
        \colhead{} & \colhead{\msun} & \colhead{\msun} & \colhead{}
    }
    \startdata
    P01 &  314 & 169 & 1.9\\ 
    P02 &  592 & 114 & 5.2\\ 
    P03 &  256 & 108 & 2.4\\ 
    P04 &  323 & 110 & 2.9\\ 
    P05 &  131 &  97 & 1.3\\ 
    P06 &   94 &  93 & 1.0\\ 
    P07 &  183 &  89 & 2.0\\ 
    P08 &   62 &  91 & 0.7\\ 
    P09 &   55 &  93 & 0.6\\ 
    P10 &   85 &  83 & 1.0\\ 
    \enddata
    \end{deluxetable}

    The magnetic field also plays an important role in shaping filamentary
    molecular clouds by providing the supporting force against the gravity or
    directing the mass accretion flow \citep{2019FrASS...6....5H}. 
    Supercritical filaments 
    are usually observed to be perpendicular to the magnetic fields 
    \citep{2017ApJ...847..114L}, whereas the striations that are attached
    to the main filaments are highly elongated along the B-field 
    \citep{2016MNRAS.461.3918H}. In the case of DR~21, the large scale
    magnetic field runs almost perpendicular to the 
    north-south main ridge consisting of DR~21 and DR21(OH) 
    \citep[e.g.][]{2009ApJ...694.1056K}. 
    The accretion flow onto the DR21(OH) through the 
    sub-filaments F3, on the other hand, has been reported being parallel to 
    the magnetic field \citep{2010A&A...520A..49S}.
    We illustrate the large scale magnetic field orientation 
    around DR~21 derived by Planck dust continuum polarization 
    data\footnote{\url{https://pla.esac.esa.int/}}
    in Figure~\ref{fig:overview}. DR21SF is mostly aligned along 
    the large scale magnetic field, and is more consistent with the
    scenario that DR21SF is overall a large scale accretion flow.
    Even though in theory, the length scale
    of ``sausage'' fragmentation is maintained with the presence of
    a magnetic field parallel to the filament \citep{1987PThPh..77..635N},
    we do not see such a fragmentation occurs in DR21SF as it does in
    many supercritical filaments.
    At the northern end of DR21SF, where the two YSO candidates are found, 
    the magnetic
    field bends to form a large angle with the spine of the filament.
    With the limited resolution of Planck data, it is difficult to tell 
    whether this alteration 
    of the magnetic field orientation is caused by the star formation
    in DR21SF or by the influence of DR~21, where strong magnetic field 
    almost perpendicular to the ridge is seen. High resolution
    observations of the dust polarization towards DR21SF
    could help us to obtain a better understanding of the stability
    status of DR21SF.

    How will such a filament dynamically evolve? 
    \citet{2018A&A...620A..62G} compared two star-forming filaments
    with similar mass and distinct central density,
    Serpens filament and Serpens south filament.
    They infer that the slightly super-critical filament (Serpens 
    filament) with lower 
    central density and weaker signature of radial collapse is on an 
    earlier evolutionary stage prior to the other's (Serpens south filament),
    which is consistent with the relatively inactive status of Serpens
    filament in star formation.  From a theoretic respect \citet{1992ApJ...388..392I} 
    proposed that the radial collapse starts in an isothermal 
    gas cylinder when the linear mass is greater than the critical value, 
    and accelerates until the central density increase and the equation 
    of state varies much from the 
    isothermal one, then the filament may finally fragment. 
    In other words, in a filament in its early evolutionary stage, 
    the radial collapse dominates over the fragmentation. 
    DR21SF shows significantly lower central density 
    compared with other star-forming filaments 
    \citep[see Table~3 in][]{2018A&A...620A..62G},
    which may indicate an early evolutionary stage.
    This is confirmed by the detection of sparse YSOs in DR21SF.
    Thus the theory by \citet{1992ApJ...388..392I} can be a possible
    explanation of the coherent profile of DR21SF.
    However, direct observational evidences of radial collapse
    are in need to verify this hypothesis in DR21SF. 

    DR21SF seems to be quiescent since we find no
    signpost for massive star formation (e.g., \ion{H}{2} regions 
    or Class II methanol masers) and see only two sites of 
    low-/intermediate-mass YSOs (P03 and P10) along the filament with no
    energetic feedbacks (e.g., shocks produced by jets or outflows). 
    Whereas the star formation sites in other 
    sub-filaments connecting to DR21(OH) are 
    more active with the detection of outflows
    (see Figure~\ref{fig:overview} for a comparison).
    There are other filamentary objects which are similar to DR21SF
    in terms of stability status. 
    In the Galactic central molecular zone (CMZ), filaments are often 
    being massive and quiescent at the same time due to the
    intensive turbulence. For example the G0.253+0.016 cloud (the ``Brick'')
    is a turbulence dominated filamentary system located in CMZ, 
    with a mass of $7.2 \times 10^4$~\msun, 
    and an apparent size of 17~pc$^2$, 
    and only two dense core candidates and one \water\ maser 
    are detected \citep[][and references therein]{2016ApJ...832..143F,
    2014ApJ...795L..25R}.  However, the physical condition of CMZ
    is apparently more extreme than that of DR21SF.
    The gas temperature is up to 100~K and one dimensional Mach number 
    in CMZ is 30 \citep{2014ApJ...795L..25R}, which are significantly
    higher than those in DR21SF. \citet{2016ApJ...832..143F} claimed
    that the solenoid driving caused by shearing motions
    in CMZ is the dominant turbulence driving model in G0.253+0.016, 
    which reduces the star formation rate compared to typical clouds
    in solar neighborhood. In our case, the strong gravitational field 
    produced by DR 21 may be a potential engine of compressing driving
    turbulence. 

  \subsection{A Scenario of Gas In-falling of DR21SF} \label{subsec:scenario}

    There are in principle three possible interpretations of the 
    radial velocity gradient along filamentary molecular clouds: 
    rotation, outflow, or in-fall.  
    The rotation scenario is unrealistic since the centrifugal force needed
    to maintain the 3.6 pc filament is too dramatic.
    Considering an outflow scenario, 
    we assume that the filament is a mass 
    ejection originating from DR~21. In 
    this case, stellar winds or jets would be the most likely 
    candidate sources for driving the outflow motions. 
    However, DR 21 is already 
    harboring one of the most luminous and largest warm (2000~K) bipolar 
    outflows in the Galaxy \citep{1986MNRAS.220..203G}. 
    There is not any star formation theory that could provide a feasible 
    interpretation of two such large outflows, one warm and bipolar and the 
    other cold and unipolar, coming from a cluster of UC \ion{H}{2} regions. 
    Since neither the rotation nor the outflow scenario appears to be feasible,
    we propose that the radial velocity gradient of DR21SF is 
    produced by an accelerating in-fall along the filament. 
    The gas flows from the south to the north along DR21SF,
    induced by the gravitational field of DR~21.
    \citet{2013A&A...553A.119A} proposed that in this 
    process the radially collapsing filaments will keep a constant width 
    along their spine. This is also in agreement with our observation.

    The system consisting of DR21SF and DR~21 seems to be
    quite similar to the ``head-tail'' system reported 
    by \citet{2002A&A...385..909T}, in which the ``head'' is an active
    massive star forming region with a high column density and the much 
    less dense ``tail'' often shows only signposts
    of low-mass star formation. \citet{2002A&A...385..909T} claimed
    that in the ``head-tail'' scenario, stellar winds from other OB stars in the 
    opposite position of the ``tail'' are likely to be the trigger of the
    formation of the cometary structure.  
    We believe in DR21SF this is not likely 
    to be the case, since the massive star-forming site to the 
    north, DR21(OH), is still at an early evolutionary stage 
    and the radiation and winds from embedded high-mass protostars not yet
    expel surrounding molecular clouds.
    Therefore they are not expected to produce a ``head-tail'' scenario
    at DR~21.  Thus the DR21SF-DR~21 system looks similar to a ``head-tail'' 
    picture, but the system is not induced by nearby OB stars, and the ``tail'' 
    is an accretion flow onto the ``head''.  

\begin{figure}
    \plotone{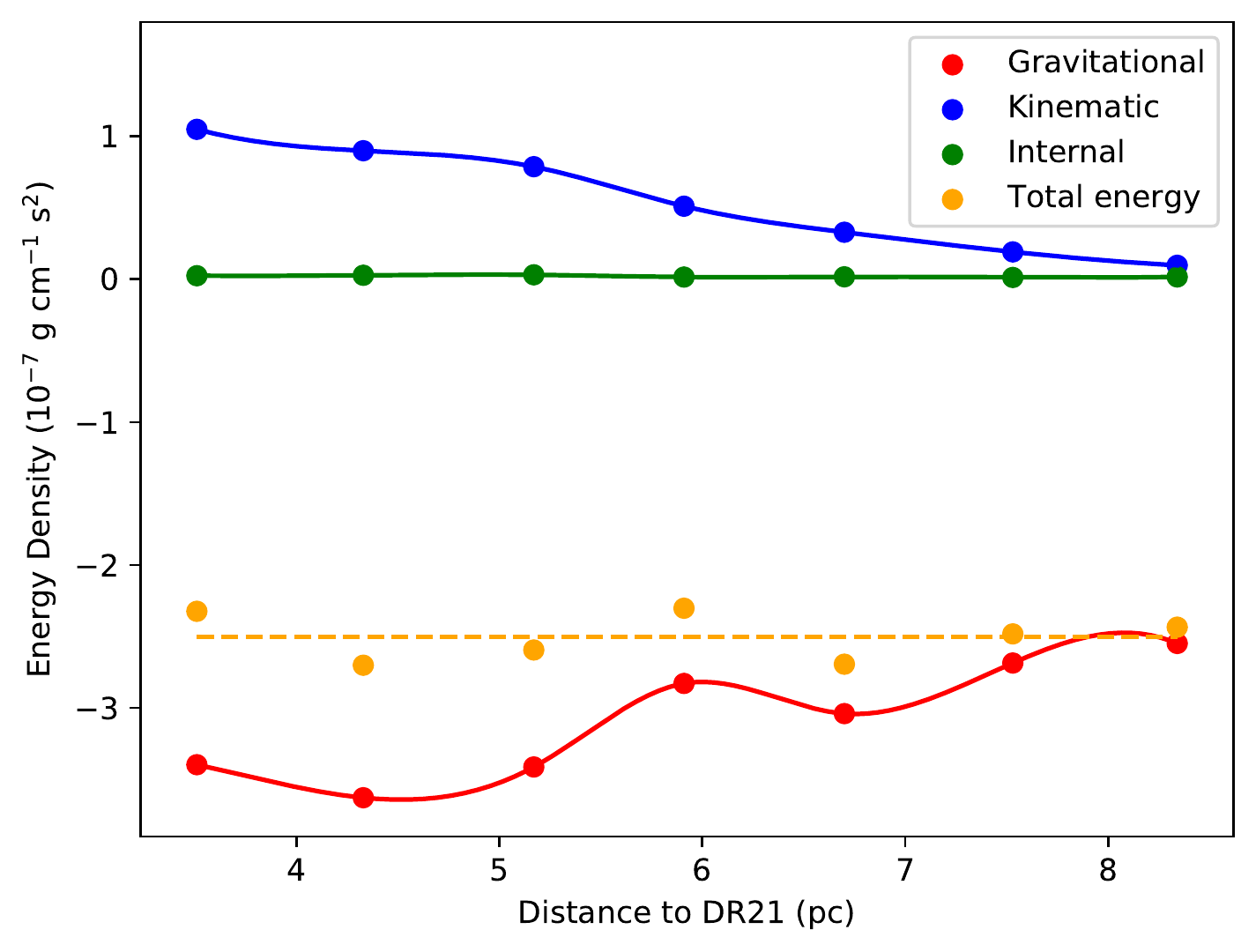}
    \caption{The gravitational energy, kinematic energy and internal
             energy of gas in an unit volume at
             P03, P04, P05, P06, P07, P08, and P09 are denoted by
             red, blue, and green dots, respectively, with spline interpolation
             to give a rough impression of how the energies are distributed
             along the DR21SF. The yellow dots illustrate the total energy
             given by the best-fit. The yellow dashed horizontal line is a 
             linear fit of the total energy showing the total energy preserves
             along the filament.
             \label{fig:energy}}
    \end{figure}

    Becaused of the conservation of the total energy during the accelerating 
    in-fall, one can estimate the
    inclination angle of the filament by analyzing its energy budget,
    assuming the a free-fall gas cylinder model. 
    Width an inclination angle $\theta$, with respective to the plane of 
    sky, the first order estimate of the 
    total energy of the in-fall gas in a unit volume reads
    \begin{equation}
    E_{\rm prot} + E_{\rm kin} + E_{\rm turb} + E_{\rm therm} = C,
    \end{equation} 
    where $E_{\rm prot} = \frac{G m_i m_{\rm DR21} \cos{\theta}}{r_i}$ 
    is the gravitational energy of gas with mass 
    $M_i$ at projected distance $r_i$ to DR~21 in the plane of sky,
    $E_{\rm kin} = m_i \frac{v_{\rm LSR}}{\sin{\theta}}^2$ is the kinetic 
    energy with relative radial velocity $v_{\rm LSR}$ in respect of the 
    southern
    end of DR21SF, and $E_{\rm turb} + E_{\rm therm} = m_i \Delta v^2$ is 
    the internal energy (including thermal and turbulent components) 
    with $\Delta v$ being the intrinsic line-width of the gas.
    The local mass is defined as $m_i = P_i \cos{\theta} / w$, 
    where $P_i$ is the local column density, and $w = 0.13$~pc is 
    the width of DR21SF.
    We again use the column density, radial velocity and intrinsic 
    line-width from P03 to P09 to fit the model. The best-fit model 
    is illustrated
    in Figure~\ref{fig:energy}, which gives an estimate of the inclination 
    angle of 18\degr.  Thus the projection effect is insignificant in
    the analysis of DR21SF. \citet{2009ApJ...694.1056K} estimated the
    inclination angle of the DR~21 main ridge is about 20\degr\ to
    the plane of the sky by fitting line-width ratio of emissions from
    neutral molecules and ions in the magnetic field. This result is
    in good agreement with ours.

    We then give an estimate of the mass in-fall rate of DR21SF 
    with the assumption of a homogeneous 
    cylinder\citep{2014A&A...561A..83P}:
    \begin{equation} 
    \dot{M} = \pi(w/2)^2\rho V = \frac{M}{L} V,
    \end{equation}
    in which $w$ is the width of the filament, $\rho$ is the average 
    density, $V$ is the in-fall velocity at the northern end of the
    filament, where we adopt the velocity of P01, $M$ is the total 
    mass, and $L$ is the length of the filament.
    The in-fall velocity can be estimated by radial velocities as
    $V_{\rm infall} = (V_{\rm P01}-V_{\rm DR21})/\sin{\theta} = 3.7$ \kms, 
    where $V_{\rm DR21} = -3.0$ \kms. 
    Thus the mass in-fall rate is
    $\dot{M} = 1.1 \times 10^{-3}$ \msunpyr. 
    Our estimate is about two orders of magnitude larger than 
    the estimate of the mass in-fall rate of SDC13 filament 
    \citep[$2.5\times 10^{-5}$ \msunpyr,][]{2014A&A...561A..83P}, 
    and is similar to that of F3 filament onto the DR21(OH) clump
    \citep[$1.9 \times 10^{-3}$ \msunpyr,][]{2010A&A...520A..49S}, which 
    is believed to contribute about one third of the total mass in-fall
    rate of the clump. Since there is no detected gas flow towards
    DR 21 along the DR21 ridge \citep{2010A&A...520A..49S}, the gas in-fall
    along the DR21SF is possibly the most important, if not only, accretion 
    flow that is feeding DR 21 massive star forming region.

\section{Summary} \label{sec:summary}
  DR21SF is a massive filament distinct from 
  the major star-forming filaments or the low-density striations associated 
  with the major filaments.  Based on the column density
  map derived from \herschel\ images, we find that DR21SF
  is a coherent filament with no clear signs of fragmentation,
  but still hosting star formation activities.
  Several low-/intermediate-mass YSOs are identified in the filament,
  with one candidate of star forming core in its extremely early evolutionary 
  stage.
  The temperature, optical depth, radial velocity, and the velocity dispersion 
  of the filament is derived by \ammonia (1, 1) and (2, 2) emission 
  line observations.
  With a size of 3.6~pc by 0.13~pc and a mass of 1048~\msun,  
  The filament has an average temperature typical of cold molecular clouds, 
  but the sites of low-/intermediate-mass YSOs show temperature excess.
  The lack of signatures of fragmentation in DR21SF, compared to other 
  filaments in the Cygnus X north filamentary network, may be due to the 
  early evolutionary stage or the unique external gravitational field 
  produced by DR 21. 

  A radial velocity gradient of 0.8~\kmsp\ is detected along the spine 
  of DR21SF.  We interpret these gradients with a scenario of an accelerating 
  gas flow in DR21SF towards DR~21, with a mass transfer rate of 
  $1.1 \times 10^{-3}$~\msunpyr.  
  We also proposed an energy-budget 
  method to estimate the inclination angle of the filament under the
  assumption of such a scenario.
  Our observations shed new lights on the filament network around DR~21,  
  and provide hints that the source of the most energetic outflow in our 
  Galaxy may have a major mass supply from a parsec scale accretion flow.

\vspace{5mm}

\acknowledgements
B.H., K.Q., Y.C., J.L., Y.W., and J.W. are supported by National Key R\&D
Program of China No.\ 2017YFA0402600. We acknowledge the support from National
Natural Science Foundation of China (NSFC) through grants No.\ U1731237, 
11473011, 11590781, 11629302, 11590780, 11590784, 12073064, and the Scientific 
Program of Shanghai Municipality (08DZ1160100).

\facilities{TianMa Radio Telescope (TMRT)}

\software{Astropy \citep{2013A&A...558A..33A}, GILDAS}

\end{document}